\documentclass[12pt,preprint2]{aastex} 
\usepackage{graphicx}

\newcommand {\halpha} {\mbox{H$\alpha$}}

\newcommand {\kms} {\mbox{~km$\,$s$^{-1}$}}

\begin{document}

\setlength{\unitlength}{1 cm}
\def\msol{\hbox{\kern 0.20em $M_\odot$}}
\def\lsol{\hbox{\kern 0.20em $L_\odot$}}
\def\rsol{\hbox{\kern 0.20em $R_\odot$}}
\def\sr{\hbox{\kern 0.20em sr}}
\def\srmu{\hbox{\kern 0.20em sr$^{-1}$}}
 
\def\g{\hbox{\kern 0.20em g}}
\def\gmu{\hbox{\kern 0.20em g$^{-1}$}}
\def\kg{\hbox{\kern 0.20em kg}}
\def\pc{\hbox{\kern 0.20em pc}}
 
\def\mum{\hbox{\kern 0.20em $\mu$m}}
\def\mumd{\hbox{\kern 0.20em $\mu$m$^{-2}$}}
\def\cm{\hbox{\kern 0.20em cm}}
\def\m{\hbox{\kern 0.20em m}}
\def\km{\hbox{\kern 0.20em km}}
\def\nm{\hbox{\kern 0.20em nm}}
 
\def\s{\hbox{\kern 0.20em s}}
\def\h{\hbox{\kern 0.20em h}}
\def\sec{\hbox{\kern 0.20em sec}}
\def\min{\hbox {\kern 0.20em min}}
\def\smu{\hbox{\kern 0.20em s$^{-1}$}}
\def\smd{\hbox{\kern 0.20em s$^{-2}$}}
\def\an{\hbox{\kern 0.20em an}}
\def\anmu{\hbox{\kern 0.20em an$^{-1}$}}
\def\deg{\hbox{\kern 0.20em $^{\rm o}$}}
\def\yr{\hbox{\kern 0.20em yr}}
\def\yrmu{\hbox{\kern 0.20em yr$^{-1}$}}
\def\Myr{\hbox{\kern 0.20em Myr}}
\def\Mymu{\hbox{\kern 0.20em Myr$^{-1}$}}
\def\K{\hbox{\kern 0.20em K}}
\def\pcmu{\hbox{\kern 0.20em pc$^{-1}$}}
\def\pcmd{\hbox{\kern 0.20em pc$^{-2}$}}
\def\pcmt{\hbox{\kern 0.20em pc$^{-3}$}}
\def\kms{\hbox{\kern 0.20em km\kern 0.20em s$^{-1}$}}
\def\kmpd{\hbox{\kern 0.20em km$^{2}$}}
\def\kpc{\hbox{\kern 0.20em kpc}}
\def\cms{\hbox{\kern 0.20em cm\kern 0.20em s$^{-1}$}}
\def\erg{\hbox{\kern 0.20em erg}}
\def\ergs{\hbox{\kern 0.20em erg}}
\def\cmpd{\hbox{\kern 0.20em cm$^2$}}
\def\cmmd{\hbox{\kern 0.20em cm$^{-2}$}}
\def\cmms{\hbox{\kern 0.20em cm$^{-6}$}}
\def\cmpt{\hbox{\kern 0.20em cm$^3$}}
\def\cmmt{\hbox{\kern 0.20em cm$^{-3}$}}
\def\mpd{\hbox{\kern 0.20em m$^2$}}
\def\mmd{\hbox{\kern 0.20em m$^{-2}$}}
\def\mpt{\hbox{\kern 0.20em m$^3$}}
\def\mmt{\hbox{\kern 0.20em m$^{-3}$}}
\def\mujy{\hbox{\kern 0.20em $\mu$Jy}}
\def\mjy{\hbox{\kern 0.20em mJy}}
\def\Mj{\hbox{\kern 0.20em MJy}}
\def\jy{\hbox{\kern 0.20em Jy}}
\def\ghz{\hbox{\kern 0.20em GHz}}
\def\srmd{\hbox{\kern 0.20em sr$^{-1}$}}
\def \cc{$\rm{cm}^{-3}$}
\def \mum{$\mu$m}

\def\G{\hbox{\kern 0.20em G}}

\def\twco{\hbox{${}^{12}$CO}}
\def\twcotwo{\hbox{${}^{12}$CO(2-1)}}
\def\twcoseven{\hbox{${}^{12}$CO(7-6)}}
\def\thco{\hbox{${}^{13}$CO}}
\def\thcotwo{\hbox{${}^{13}$CO(2-1)}}
\def\thcoone{\hbox{${}^{13}$CO(1-0)}}
\def\ceio{\hbox{C${}^{18}$O}}
\def\ceiotwo{\hbox{C${}^{18}$O(2-1)}}
\def\ceioone{\hbox{C${}^{18}$O(1-0)}}
\def\cs{\hbox{CS}}
\def\csthree{\hbox{CS(3-2)}}
\def\cstwo{\hbox{CS(2-1)}}
\def\csfive{\hbox{CS(5-4)}}
\def\cts{\hbox{C${}^{34}$S}}
\def\ctsthree{\hbox{C${}^{34}$S(3-2)}}
\def\ctstwo{\hbox{C${}^{34}$S(2-1)}}
\def\htwo{\hbox{H${}_2$}}
\def\h13cop{\hbox{H$^{13}$CO$^{+}$}}
\def\halpha{\hbox{H$\alpha$ }}
\def\hcop{\hbox{HCO$^{+}$}}
\def\Sp{\hbox{S{\small II}}}
\newcommand{\Hp}{\hbox{H{\small II}}}
\newcommand{\Cp}{\hbox{[C{\small II}]}}
\newcommand{\Opp}{\hbox{[O{\small III}]}}
\newcommand{\Op}{\hbox{[O{\small I}]}}
\newcommand{\Np}{\hbox{[N{\small II}]}}
\newcommand{\Nep}{\hbox{[Ne{\small II}]}}
\newcommand{\Nepp}{\hbox{[Ne{\small III}]}}
\newcommand{\Npp}{\hbox{[N{\small III}]}}                                      
\newcommand{\Sip}{\hbox{[Si{\small II}]}}                                      
\newcommand{\jonetozero}{\hbox{$J=1\rightarrow 0$}}
\newcommand{\jtwotoone}{\hbox{$J=2\rightarrow 1$}}
\newcommand{\jthreetotwo}{\hbox{$J=3\rightarrow 2$}}
\newcommand{\jfourtothree}{\hbox{$J=4\rightarrow 3$}} 
\newcommand{\jfivetofour}{\hbox{$J=5\rightarrow 4$}}                  
\newcommand{\jsixtofive}{\hbox{$J=6\rightarrow 5$}}      
\newcommand{\jseventosix}{\hbox{$J=7\rightarrow 6$}}

\title{Anatomy of HH~111 from CO observations~: a bow shock driven molecular 
outflow}

\shorttitle{The HH~111~Jet}
\shortauthors{Lefloch et al.}

\author {Bertrand Lefloch \altaffilmark{1},
Jos\'e Cernicharo \altaffilmark{2},
     Bo Reipurth \altaffilmark{3},
Juan Ramon Pardo\altaffilmark{2}
and        Roberto Neri \altaffilmark{4}
}

\altaffiltext{1} {Laboratoire d'Astrophysique de l'Observatoire de 
Grenoble, BP 53, 38041 Grenoble cedex, France}

\altaffiltext{2} {Instituto de Estructura de la Materia, Dpto 
F\'\i sica Molecular, Serrano 123, 28006 Madrid, Spain}

\altaffiltext{3} {Institute for Astronomy, University of Hawaii, 
640 N. Aohoku Place, Hilo, HI 96720, USA}

\altaffiltext{4} {IRAM, Domaine Universitaire, 300 rue de la Piscine,
38406 St Martin d'H\`eres, France}


\begin{abstract}
We present single-dish and interferometric millimeter line observations of the 
HH~111 outflow and its driving source. 
The physical conditions of the protostellar core have been determined from the 
emission of the millimeter line emission of CO and its 
isotopomers and CS with the IRAM 30m telescope,
and the CO \jseventosix\ line with the Caltech Submm Observatory. The 
molecular gas emission
reveals a small condensation of cold ($\rm T= 20-25\K$) and dense gas 
($n(\htwo)= 3\times 10^5\cmmt$). 
The low-velocity outflowing gas has been mapped with the IRAM Plateau de
Bure interferometer. The cold gas 
is distributed in a hollow cylinder surrounding the optical jet.  The 
formation of this cavity and its kinematics are well accounted 
for in the frame of outflow gas entrainment by jet bow shocks. 
Evidence of gas acceleration is found along the cavity walls, 
correlated with the presence of optical bow shocks.
The cavity has been expanding with a mean 
velocity of $4\kms$ on a timescale of $8700\yr$, similar to the dynamical age 
of the optical jet. The separation of the inner walls reaches 
$8\arcsec-10\arcsec$, which matches the transverse size of the wings in the 
bow shock. 
CSO observations of the \jseventosix\ line show evidence of a high-velocity 
and 
hot gas component ($\rm T=300-1000\K$) with a low filling factor. This
emission probably arises from shocked molecular gas in the jet. Observations
of the $\rm ^3P_2-^3P_1$ [CI] line are consistent with C-type 
non-dissociative shocks.
Mapping of the high-velocity molecular bullets B1-B3, located beyond 
the optical jet, with the IRAM PdBI
reveals small structures of $3\arcsec \times 7\arcsec$ flattened 
perpendicular to the flow direction. 
They are made of cold ($\rm T\sim 30\K$), moderate density gas 
($\rm n(\htwo)= (0.5-1.0)\times 10^4\cmmt$). We find evidence that 
the bullets are expanding into the low-density surrounding medium.
Their properties are consistent with their being shocked gas 
knots resulting from past time-variable ejections in the jet. 
\end{abstract}  


\section{Introduction}

Herbig-Haro jets are manifestations of the accretion processes that
are involved in the birth of stars, and as such represent a fossil
record of the recent activity of their driving sources. The study of
jets thus enables us to understand the past history of young stars as
they build up their mass. As these jets penetrate their ambient
medium, they transfer momentum and accelerate their surroundings and
it appears that it is this interaction which causes the ubiquitous
molecular outflows. For a recent review on these issues, see Reipurth
\& Bally (2001).

One of the finest HH jets known is the HH 111 jet, located in the
L1617 cloud in Orion (Reipurth 1989). It has a very highly collimated
body, with a large number of individual knots, and has large proper
motions (300 to $600\kms$). The jet flows at an angle of only
10$^o$ to the plane of the sky (Reipurth, Raga, \& Heathcote
1992). Beyond the collimated jet the flow continues, and it is in fact
only a minor part of a truly gigantic bipolar HH complex stretching over 
7.7 pc (Reipurth, Bally, \& Devine 1997a). Optical spectra are
discussed by Morse et al. (1993a,b) and Noriega-Crespo et
al. (1993). Near-infrared observations have revealed a remarkable
symmetry between the optical jet and a near-infrared counterjet, and
that a second bipolar flow, HH~121, emerges from the source,
suggesting that it is a binary (Gredel \& Reipurth 1993, 1994; Davis,
Mundt, \& Eisl\"offel 1994, Coppin et al. 1998, Reipurth et
al. 2000). The optical jet is co-axial with a major well collimated
molecular outflow (Reipurth \& Olberg 1991, Cernicharo \& Reipurth
1996, Nagar et al. 1997, Lee et al. 2000).  Beyond the visible jet,
Cernicharo \& Reipurth (1996) (henceforth CR) found a set of three high
velocity CO bullets, which apparently represent working surfaces that
no longer produce optical shock emission.  The jet is driven by IRAS
05491+0247, a Class~I source with a luminosity of about 25 L$_\odot$,
surrounded by cold dust and gas (Reipurth et al. 1993, Stapelfeldt
\& Scoville 1993, Yang et al. 1997, Dent, Matthews, Ward-Thompson
1998). A small 3.6 cm radio continuum jet was detected around the
source by Rodr\'\i guez \& Reipurth (1994). Recently, infrared images
obtained with {\em HST} have revealed that another source is located
only 3 arcsec from the main one, which itself is a binary judging
from the quadrupolar outflows; so altogether it appears that the
driving source is part of a hierarchical triple system (Reipurth et
al. 1999). In fact, the entire giant jet flow may be a result of the
dynamical interactions between the three members of this triple system
(Reipurth 2000). 

The short blue lobe found by Reipurth \& Olberg (1991) appears to be  
a highly collimated molecular jet about $90\arcsec$ long 
with a collimation factor of 9 (see their Figs.~2-3-4) in the 
$12\arcsec$ data of CR. The blue lobe ends 
at the bow shock P and the optical knots Q,R,S. Nagar et al. (1997) showed
an interferometric map at $6\arcsec$ resolution of the $\jonetozero$ line, 
which 
allowed them to resolve the structure of the low-velocity gas around the 
optical jet.  The main features of the ionized and molecular gas in HH~111 are 
summarized in Fig.~1.

In this paper we present single dish and high resolution interferometric 
observations of the body of HH~111, the protostellar core and the high
velocity CO bullets.  We characterize the gas kinematics and the physical 
conditions in the gas 
and report the detection of a hot temperature component in the outflow. 
We discuss these results in relation to current
theories on the coupling of HH jets and molecular outflows.
  
We find that the 
molecular gas entrainment in the outflow is well accounted for in the frame
of jet bow shock driven outflows, as modelled by Raga \& Cabrit (1993) 
(henceforth RC93).
Preliminary results were reported by Reipurth \& Cernicharo
(1995) and Cernicharo, Neri \& Reipurth (1997). 

\section{Observations}

\begin{figure*}
  \begin{center}
    \leavevmode
\caption{Map of the CO \jonetozero\ outflow emission integrated 
between $+0.25\kms$ and $+7.3\kms$ obtained with the IRAM Plateau de Bure 
interferometer. We have indicated the main features detected in the molecular 
gas (the cavity and the high velocity bullets) gas and in the ionized gas 
(the optical jet, as observed with HST, the shocked knots  
and the large bow shock V). 
A dotted line traces the main axis of the jet. 
}
  \end{center}
\label{plot_velmap}
\end{figure*}

\subsection{Single-Dish Data}

The central source and the HH~111 outflow were observed in the millimeter 
lines of CS and the CO isotopomers with the IRAM-30m telescope in 
1990 and 1994. The CO observations have already been presented in CR. 
For these observations, we used as spectrometer an 
autocorrelator with 2048 channels and spectral resolution of 37 kHz. The 
spectral resolution was degraded to $\approx 0.2\kms$. The results 
are expressed in units of main-beam brightness temperature. The main-beam 
efficiency was measured to be 0.63 and 0.57 at the frequency of the CS 
\jtwotoone\ and \jthreetotwo\ transitions, respectively. 

The CO \jseventosix\ line at 806.651708~GHz was observed towards the central 
source and bullet B3 with the Caltech Submm Observatory (CSO) in December 
2005. The receiver was tuned in double-side band and the 
system temperature ranged between 5700 and 7500 K. This allowed us to 
observe simultaneously the $\rm ^3P_2-^3P_1$ [CI] lines at 809341.97~GHz 
in the image band. An AOS was used as 
spectrometer, which provided a kinematical resolution of $0.24\kms$. At this 
frequency, the beam size of the telescope is about $12\arcsec$ and the beam 
efficiency is 0.28. The calibration was checked towards Ori IRc2 in Orion A.

\subsection{Interferometric data}

The CO \jonetozero\ emission along the HH~111 outflow was observed with the 
IRAM Plateau de Bure interferometer (PdBI). A mosaic of 5 fields was 
performed; 
the field centers were located at the offset positions (+10\arcsec,0), 
(-10\arcsec,+3\arcsec), (-30\arcsec,+6\arcsec),(-50\arcsec,+9\arcsec) and 
(-70\arcsec,+12\arcsec)  with respect to the source respectively~:
$05^h 51^m 46.2^s$ $+02^{\circ} 48' 29.5\arcsec$ (J2000). 
0528+134 and 3C120 were used as phase calibrators. 
An autocorrelator was used as spectrometer, with an element of resolution
of 20~kHz, corresponding to a kinematic resolution of $\approx 0.06\kms$
at the CO \jonetozero\ frequency.
The IRAM 30m data of the region obtained by CR 
were used to provide the short spacings and combined with the PdBI 
visibilities. The combined map is shown in Fig.~5. 
The synthetic beam  is $3.79\arcsec \times 2.43\arcsec$ at 2.6~mm and makes 
a P.A. of $-4^o$ deg with respect to North. 

The CO \jtwotoone\ emission of the high-velocity bullets B1, B2 and B3 
discovered outside the molecular cloud was observed with the IRAM PdBI in 
1996. Three fields were observed centered respectively at  
$05^h 51^m 34.65^s$ $+02^{\circ} 48' 50.50\arcsec$ (B1), 
$05^h 51^m 31.12^s$ $+02^{\circ} 48' 57.50\arcsec$ (B2),
$05^h 51^m 28.06^s$ $+02^{\circ} 49' 08.02\arcsec$ (B3). The same phase 
calibrators as before were used. 
The resolution of the autocorrelator was set to 160~kHz, which provided 
a kinematic resolution of $\approx 0.16\kms$ at the CO \jtwotoone\ frequency.
The synthetic beam is $1.96\arcsec \times 1.26\arcsec$ at 1.3~mm and makes 
a P.A. of $-192^o$ deg with respect to North.

To estimate the fraction of flux missed by the interferometer, we have 
compared the flux measured at the IRAM 30m telescope with the flux recovered 
by the interferometer towards bullet B2 in the velocity interval of maximum 
intensity intensity.  At the IRAM 30m, the line intensity peak is $0.43\K$, 
which corresponds to a total flux of $5.1\jy$, after correcting for the 
source convolution with the beam and taking into account the antenna 
efficiency of 10Jy/K at 1.3mm. 
At the same velocity ($-36.8\kms$), the flux integrated over the reconstructed 
source was  estimated with the ``FLUX'' procedure and found to be 
$4.95\jy$. We conclude that almost all the flux has been recovered by the 
interferometer. 

\section{The source}

\begin{figure}
  \begin{center}
    \leavevmode
    \includegraphics[width=\hsize]{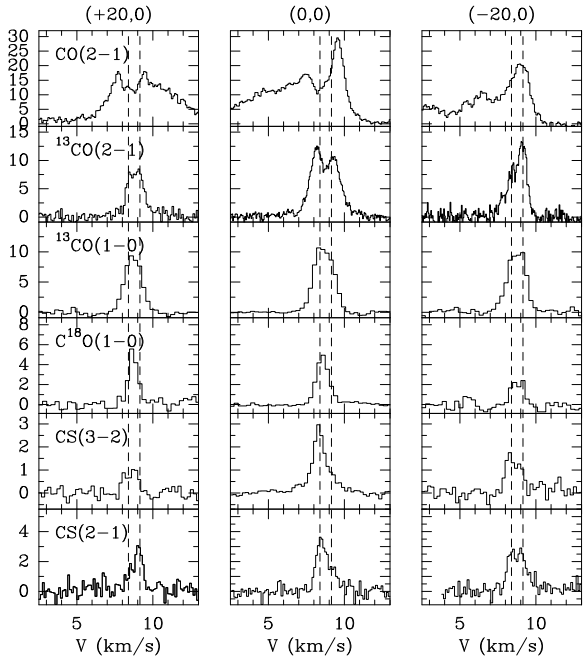}
\caption{({\em bottom})~
Molecular emission of CO, $\thco$, \ceio\, CS and DCO$^{+}$ towards selected
positions along the optical jet of HH111, indicated above each upper panel. 
The dashed lines mark the 
velocity of the protostellar core ($v= 8.4\kms$) and 
the cavity wall ($v=9.2\kms$). }
  \end{center}
\label{plot_velmap}
\end{figure}

Figure~2 shows the emission of  the millimeter lines of CS  and the CO 
isotopomers detected towards the VLA source and two nearby positions offset
by $20\arcsec$ in right ascension. In the direction of the VLA source, 
double-peaked profiles are detected in all lines of various opacities, but the 
$\ceio$ $\jonetozero$ and $\thco$ $\jonetozero$ 
(see, e.g., CS $\jtwotoone$ and $\thco$ $\jtwotoone$).
In particular, the maximum of the singly peaked \thco\ \jonetozero\ 
coincides with the $8.4\kms$ peak of the \jtwotoone\ transition. 
This allows us to exclude self-absorption as origin of the double peaked 
profiles detected. 

The $^{13}$CO interferometric map of HH~111 by Stapelfeldt \& Scoville (1993)
indicates a North-South velocity gradient in the direction of the core
where the powering source of HH~111 has been formed.  The CS map
of Yang et al. (1997) taken with the Nobeyama interferometer also
suggests such a velocity gradient. It has been interpreted by Yang et
al. (1997) as due to infalling gas around the exciting source of
HH~111. However, the fact that the emission from the CS lines show
wings associated with the molecular outflow renders
difficult the interpretation of the data. Indeed, the gas kinematics 
is affected by several different physical processes, as evidenced by the 
presence of a well defined cavity (see below Sect.~4.3; also CR and 
Nagar et al. 1997).

We do not detect any spatial shift in the velocity centroid of the molecular
emission in  the direction perpendicular to the outflow. 
Hence, the double peaked line profiles are not tracing the 
collapsing gas of the protostellar envelope, but most likely
the interaction of the jet with the ambient gas, constituting the rear
part of an expanding cavity around the optical jet. This point is adressed 
more thoroughly in  Sect.~4. 

The $\ceio$ $\jonetozero$ transition traces the emission of 
the parental core at $8.4\kms$. The $\ceio$ and $\thco$ \jonetozero\ 
were observed simultaneously by the same receiver (they lie in the 
same IF band), 
which offers the advantage of eliminating the uncertainties in the pointing 
and the relative calibration. 
Towards the source, we measure main-beam brightness temperatures 
$\rm T^{18}= 4.9\K$ and $\rm T^{13}= 10.6\K$ 
for the $\ceio$ and the $\thco$ transitions, respectively. Both 
transitions are thermalized at the density of the core. 
Assuming a standard relative abundance ratio $\rm [^{13}CO]/[C^{18}O]= 8$, 
we derive an opacity $\tau_{10}^{18}= 0.56$ for the \ceio\ line. 
Analysis of the \ceio\ transition in the Large-Velocity Gradient 
approximation yields a gas column density 
$\rm N(\ceio)= 6.5\times 10^{15}\cmmd$ and a temperature of $15\K$. 
Hence, the total gas column density of the core is 
$\rm N(\htwo)= 3.3\times 10^{22}\cmmd$. 

The CO $\jseventosix$ to 
$\jtwotoone$ intensity line ratio measured  at ambient velocities, 
between +3.8 and $+11.5\kms$ (see Sect. 4.1), is $\approx 0.3$, 
as expected for gas at $20-25\K$ and densities of $10^4-10^5\cmmt$. 
We therefore adopt a value of $20\K$ in what follows. 

The dense gas tracers (CS) peak at $v_{lsr}= 8.4\kms$ in the 
central core region. The lines are rather bright ($\rm T_{mb}= 3\K$)
and narrow ($\Delta v$ less than $1\kms$). Small wings are detected 
between 6.5 and $11\kms$ (Fig.~2). The emission draws a 
round condensation of $26\arcsec$ diameter (beam-deconvolved) or $0.06\pc$.
Analysis of the CS \jthreetotwo\ and \jtwotoone\ transitions in the 
Large-Velocity Gradient approximation yields typical densities of 
$\rm n(\htwo)= 3\times 10^5\cmmt$ and column densities 
$\rm N(CS)= 1.3\times 10^{13}\cmmd$ at the brightness peak.
The lines are optically thin with $\tau^{21}= 0.44$ and $\tau^{32}= 0.77$. 

Emission in the $\rm ^3P_2-^3P_1$~[CI] line was detected 
at the velocity of the ambient gas $\simeq 9\kms$ 
(Fig.~3; note that the [CI] line is in the image band).  A gaussian 
fit to this emission yields an integrated flux of $20\K\kms$,  peak intensity
of $1.5\K$ and a line width of $12\kms$. Such a broad line width implies 
that the [CI] emission arises from the outflow and not only from the ambient 
gas. This suggests a gas column density of 
$1.0\times 10^{18}\cmmd$ for a  temperature of $20\K$. 
The ratio of atomic carbon to carbon monoxide column densities is therefore 
$\rm [CI]/CO \simeq 0.3$, a typical value measured in the Orion molecular 
cloud.

\section{The outflow}

\subsection{Hot gas near the source}

\begin{figure}
  \begin{center}
    \leavevmode
    \includegraphics[height=7cm,angle=-90]{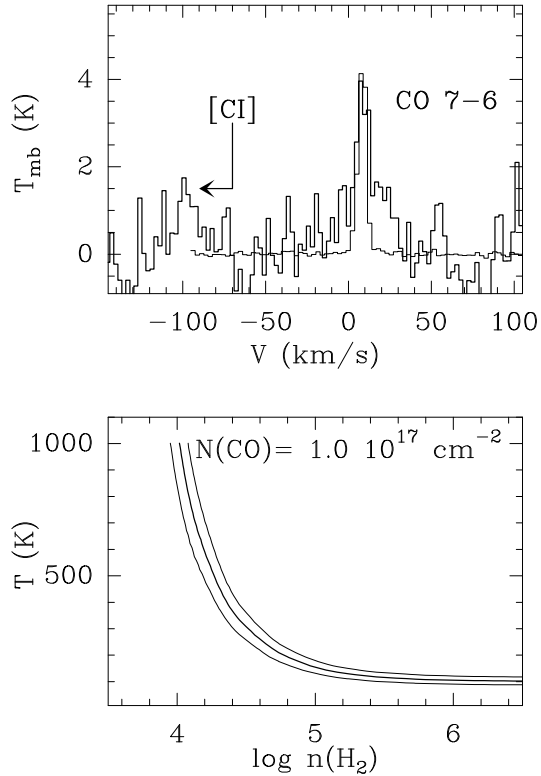}
\caption{ {\em (top)}. 
CO \jseventosix\ observed towards the central source (thick). 
We have superposed (thin) the \jtwotoone\ line emission detected with the 
IRAM 30m telescope. 
{\em (bottom)}~Variations of the 7-6/2-1 ratio R as a function 
of the kinetic temperature and density.  Calculations were done for a 
CO column density of $1.0\times 10^{17}\cmmd$ in the LVG approximation. 
The contours at (0.8,1.0,1.2) times the measured ratio R= 3.0 are drawn.  
}
  \end{center}
\label{plot_bullet}
\end{figure}

No evidence of high-velocity wings was found towards the central
source in the low \jonetozero\ and \jtwotoone\ CO line profiles 
obtained with the IRAM telescope (CR; see also Fig.~3). CR pointed
out that in case the outflowing gas would have a high excitation temperature, 
much above the 15-20~K estimated from the lowest two transitions,  
the opacity of the latter would be very low,  making them
extremely difficult to detect. 

Hatchell et al. (1999) showed evidence
for warm gas in the range $80-150\K$ in the HH~111 high-velocity bullets 
from observations in the CO \jfourtothree\ and \jthreetotwo\ transitions. 
We have observed the central source in the high 
\jseventosix\ CO transition at the CSO; the upper energy level 
lies at $155\K$ above the ground state, and is a much more sensitive probe of 
warm gas. 

The final spectrum is displayed in Fig.~3. The kinematical resolution was 
degraded to $2.4\kms$ to improve the signal-to-noise ratio. The rms is 
$0.20\K$ per velocity channel. We  have superposed the 
\jtwotoone\ spectrum obtained with the IRAM 30m 
telescope, scaled to match the brightness peak of the \jseventosix\ line.
The HPBW of both telescopes  are similar at the concerned frequencies.
We detect high-velocity wings both at blue and 
redshifted velocities, with a signal:noise of $4\sigma$ and above $5\sigma$, 
respectively. Emission is detected at velocities larger than $30\kms$ with 
respect to the source. These wings are barely seen in the \jtwotoone\ line. 
The \jseventosix\ line emission can therefore be divided in 3 velocity 
intervals~: 
the blueshifted wing ($-$40 to $+3.8\kms$), the ``ambient'' gas ($+$3.8 to 
$+13\kms$), and the redshifted wing (+13 to $+40\kms$). 

As illustrated in Fig.~3, the relative intensity of the CO \jseventosix\ with 
respect to the \jtwotoone\ is much larger  in the wings than in the 
ambient gas, which indicates highly different physical conditions, in 
particular a much higher  temperature.  
We have estimated the physical conditions in the high-velocity wings 
from modelling both lines in the Large-Velocity 
Gradient approach. We have adopted the CO collisional coefficients determined
by Flower (2001) for CO - ortho-\htwo\ collisions in the range $5\K$-$400\K$. 
For temperatures beyond $400\K$, the collisional coefficients are 
extrapolated adopting a temperature dependence of $\sqrt(T/400\K)$. 
The signal:noise ratio in the wings is not high enough to study the excitation 
conditions as a function of velocity. Therefore, we consider the 
emission integrated over the velocity interval assigned to each wing. 
We obtain a similar $\jseventosix / \jtwotoone $ line ratio $\approx 3$. 
This value is probably a lower limit as the filling factor of the 
\jseventosix\ is probably less than that of the  $\jtwotoone$. 
The LVG calculations show that, in order to account for 
a $\jseventosix / \jtwotoone $ line ratio of $\approx 3$, the kinetic 
temperature has to be above $100\K$ {\em at least}, and the density larger 
than a few $10^3\cmmt$ (see Fig.~3). In this range of density and column
density, the \jtwotoone\ transition is thermalized but the 
$\jseventosix$ is not;
the \jtwotoone\ is optically thin while the \jseventosix\ is optically thick. 

The velocity of this hot gas is very large, up to $150\kms$ when 
deprojected. If the emission arises from entrained gas
then the density of the hot gas is probably much less 
than in the cold low-velocity gas, where we estimate 
$\rm n(\htwo)= 7\times 10^4\cmmt$ 
(see Sect.~4.2). The density of the hot CO gas is at most a few times 
$10^4\cmmt$, 
which implies a temperature of about $300\K$ up to $1000\K$ or more. 
Spectroscopic diagnostics in the optical range indicate typical densities 
of $7000\cmmt$ in the jet (Hartigan et al. 1994).

The CO column density of entrained gas in each wing lies 
in the range $(0.3-1.0)\times 10^{17}\cmmd$.
The mass of warm \htwo\ is about $(0.3-1.0)\times 10^{-2}\msol$ 
per beam, and the associated momentum is $0.3-1.0\msol\kms$, {\em assuming a 
filling factor of 1}.  
Integrated over the cavity, the mass of warm gas would amount to 
$0.02-0.07\msol$
and the velocity-deprojected momentum would be $\simeq 2-7\msol\kms$. 
The overall momentum carried away in the hot CO gas would be a factor of 
10 larger 
than in the low-velocity cold gas and the high-velocity molecular jet 
(Nagar et al. 1997; CR). We conclude that the filling factor of the hot 
component is probably much smaller than 1. For a filling factor of 0.1, the 
same amount of momentum 
is carried away by the three components (molecular jet, high-velocity wings, 
low-velocity ``cavity''). As a consequence, 
the hot CO emission is to arise from a narrow region
in/around the optical jet, and the size of the emitting region is a few 
arcsec. However, 
we can not exclude that the emission arises from the whole cavity. If so, 
Kelvin-Helmholtz instabilities should grow along the jet. This has not been 
observed by HST. As discussed by Reipurth et al. (1997b; hereafter Rei97) 
the optical jet propagates into a very-low density medium 
($\rm n\sim 350\cmmt$). 
Therefore, we favor the hypothesis that the hot CO gas emission comes from 
the jet itself. 

The spectra obtained by CR in the CO 
\jtwotoone\ line further West in the flow, towards optical knots P to S, 
show similar line wing profiles, typically $30\kms$ wide, with  maximum 
intensity at the ambient gas velocity, {\em plus} the signature of the 
molecular jet detected as a $5-8\kms$ wide isolated velocity component 
centered at $\approx -60\kms$ (see panels 
{\it a} and {\it b} in Fig.~3). Assuming similar excitation conditions
in the jet,  i.e. a $\jseventosix / \jtwotoone$ line intensity ratio of 3, 
we would expect a  \jseventosix\ line intensity of $\simeq 0.4\K$,  which is 
below the detection limit allowed by our data.

\begin{figure}
  \begin{center}
    \leavevmode
    \includegraphics[width=5cm]{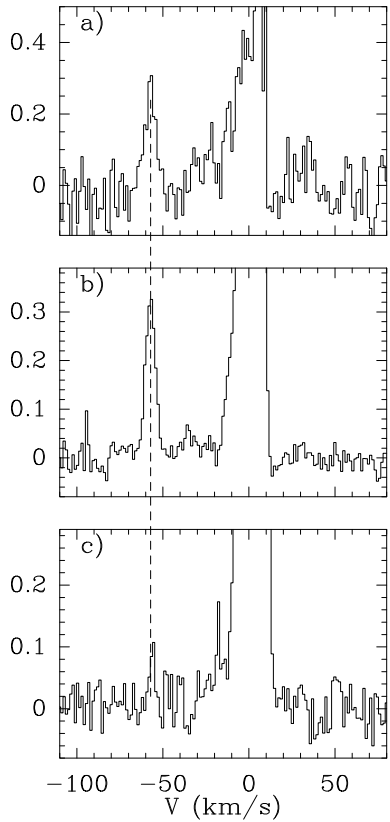}
\caption{High-velocity CO outflow emission detected with the 
IRAM 30m telescope beyond the cavity, at $96\arcsec$ from the source (a), 
near optical knots P-S (b) and towards the cavity, within $70\arcsec$ 
from the source (c). 
}
\end{center}
\end{figure}

When averaging the \jtwotoone\ line profiles over the low-velocity outflow 
emission region, between the source and offset $-70\arcsec$, a weak 
high-velocity component is detected at the $4\sigma$ level, which peaks at 
the same velocity as in the jet (see panel {\it c} in Fig.~4). Its narrow 
line width ($\Delta v \simeq 3\kms$) is consistent with a motion parallel 
to the optical jet (a spectrum with $13\kms$ kinematical
resolution is displayed in Fig.~1 of CR). Assuming a minimum excitation 
temperature of $100\K$, the mass
of this component is $4\times 10^{-4}\msol$ and the momentum carried away 
is about $0.2\msol\kms$, after correcting for velocity projection effects. 

High signal:noise \jtwotoone\ data show that the velocity distribution
between the molecular jet and the high-velocity wings is continuous (see e.g. 
the spectrum obtained near knot S  in Fig.~4a). This 
supports the idea that the high-velocity wings consist of material 
entrained by/with the jet. 
ISO observations of molecular outflows have detected the emission of 
rotational lines of CO, $\rm H_2O$ and \htwo\ at temperatures between
500 and $1200\K$ (Nisini et al. 2000; Lefloch et al. 2003). In the case 
of L1448-mm and HH~2, this hot gas emission appears to arise from extreme 
high-velocity clumps of $2\arcsec$ size along the molecular jet. 

As a conclusion, the hot CO high-velocity emission detected in HH~111
probably traces gas that has been accelerated by  internal shocks along 
the jet. 

\subsection{The physical conditions}

The single-dish data allow to determine the physical conditions in the 
outflowing gas.
The CO $\jtwotoone$ and $\jonetozero$ line intensity both peak at $+9.7 \kms$ 
(Fig.~2). The gas at this velocity extends along
the optical jet. The emission maxima are at positions
$-$10 to $-$20 arcsec away from the exciting source of the HH~111 system; it 
appears 
as bright as the main core in the lines of CO and $\thco$. 
In the \ceio\ \jonetozero\ line, a secondary component at 
$+9\kms$ is detected from $-$30 arcsec to $+$60 arcsec from the exciting 
source 
along the outflow.  A weak counterpart is detected at $7.7\kms$ about  
$50\arcsec$ from the source, which coincides with the red outflow wing. 
The parameters of the secondary component were obtained from a gaussian fit 
to the \ceio\ lines,  
which yields a line width of $0.55\kms$, a brightness temperature of $1.4\K$, 
and an emission peak at $+9.2\kms$. This implies a column density of 
$\approx 2.5\times 10^{21}\cmmd$ provided the 
kinetic temperature is similar to that of the main core, as discussed above.
We have estimated the density in the outflow wings from studying the 
secondary component in the CS lines. After removing the contribution of 
the main outflow component to the spectra, we have determined from a gaussian 
fit main-beam
peak temperatures of $1.4\K$ and $0.75\K$ for the \jtwotoone\ and 
\jthreetotwo\ lines from a gaussian fit. At a 
temperature of $20-25\K$, this implies a gas density 
$\rm n(\htwo)= 7\times 10^4\cmmt$, and a CS column density 
$\rm N(CS)= 7\times 10^{12}\cmmd$.

\begin{figure*}
  \begin{center}
    \leavevmode
\caption{
Interferometric map of the CO\jonetozero\ emission in HH~111. 
The optical HH~jet, as observed with HST, is drawn in white. 
The synthetic beam ($2.5\arcsec \times 3.5\arcsec$ HPBW) is indicated 
in the lower-right panel. First contour and contour interval in the various
panels~: $0.6\K$ in the first column; 0.6 and $0.9\K$, respectively, in the 
second column; $0.6\K$ for the panels at 7.7 and $7.2\kms$; 0.6 and 1.0 for 
the panel at $6.4\kms$; $0.4\K$ for the panels at 5.6 and $4.9\kms$ (third 
column); $0.4\K$ in the fourth column.
}
  \end{center}
\label{plot_bullet}
\end{figure*}

\subsection{The PdBI observations}

\begin{figure*}
  \begin{center}
    \leavevmode
\caption{
{\em (top)}~Map of the CO \jonetozero\ emission integrated between $+0.25$ 
and $+7.3\kms$.
The position of optical knots along the jet is marked with squares.
The white line fits the cavity of gas swept-up in the propagation of 
bow shock V. 
Two dashed lines mark the opening of the cavity inner walls. The synthesized 
beam is drawn in the lower left corner. Contours range from 
0.2 to 1.1 by step of 0.1 times 7.1 Jy/beam km/s. {\em (bottom)}~Magnified 
view of the CO cavity.}  
  \end{center}
\end{figure*}

We present in Fig.~5 the velocity channel maps of the CO \jonetozero\ emission 
between 0 and $+13\kms$, 
obtained at the PdBI. The maps result from a mosaic of 5 overlapping fields 
covering $120\arcsec \times 40 \arcsec$. Zero spacing data from  the 
IRAM 30-m telescope has 
been merged with the PdBI data. The HPBW is $2.5\arcsec \times 3.5\arcsec$  
and is indicated in the 
lower-right panel. These observations 
have a better angular resolution ($\approx$ a factor of 2), sensitivity
and {\it uv} coverage than the map of Nagar et al. (1997).  We have 
superposed in 
white the emission of the optical jet, as observed with HST in 
$\rm H\alpha$ (Rei97).  
The data confirms the presence of a hollow tubular structure with an outer 
diameter 
of $30\arcsec$ ($2.1\times 10^{17}\cm$), as previously proposed by CR and 
Nagar et al. (1997). Within $30\arcsec$ to the source, the geometry of the 
cavity is  well fitted by a parabolic law. However, 
further away from the source, the gas distribution is better described 
by a hollow conical cylinder (Fig.~6) 

The front and rear sides of the cavity are detected 
between  +2 to $+8\kms$ and +8.5 to $+12\km$, respectively, hence at 
velocities much lower than the CO outflow ($-60\kms$, CR). Line widths 
in the blueshifted gas are rather large, with $\Delta v \approx 4\kms$ at 
half-power. The walls of the cavity are seen particularly prominent at around 
$+6\kms$; they present a wider separation as the distance from the source
and the velocity increase (see Fig.~6). HST observations in the optical 
[SII] line 
indicate that the internal bow shock wings of the jet have 
a typical diameter of $5\arcsec$ (see Figs.~7-11 in Rei97) so that they
fill only partly the molecular gas cavity.  The walls extend as far as 
$60\arcsec$ away from the source ($4.1\times 10^{17}\cm$), beyond the HH jet. 
At $-70\arcsec$, we measure an inner separation of $8\arcsec$
between the walls, which corresponds to an opening angle of $5\deg$. This 
separation is also the distance between the wings of the bow shock observed 
in the jet at this position. 
Emission at extreme velocities of $+2.5$ and $+10.5\kms$ is detected 
along the jet main axis, which provides a measurement of the expansion of 
the cavity around 
the jet. We derive an average radial expansion velocity of $\approx 4\kms$. 
This implies a kinematical age of $8700\yr$ for the outermost gas layers in 
the expanding cavity, at $15\arcsec$ away from the optical jet axis. This is 
similar
to the kinematical age of the optical jet (CR).

\begin{figure*}
  \begin{center}
    \leavevmode
\caption{ Velocity-position diagram of the CO \jonetozero\ emission along cuts 
parallel to the HH jet. Positive (negative) cuts are made north (south) 
of the jet axis. Distance to the jet (in arcsec) is indicated in 
the upper left corner of each panel. First contour and contour interval
is 0.5 Jy/beam (3.7~K). In the cut at $-0.4\arcsec$ from the jet axis,
we have indicated with thick white lines the front and rear cavity walls. 
In thick black dashed is marked the direction of the secondary CO component. 
}
  \end{center}
\label{plot_bullet}
\end{figure*}

The velocity distribution on the rear side of the cavity provides strong 
constraints on its geometry. The emission of the rear wall of the cavity is 
detected at about $9\kms$, West of the source, where the outflow propagates at 
{\em blueshifted} velocities (see panel at offset position ($-$20,0) 
in Fig.~2).
The emission of the rear wall is characterized by relatively narrow 
line widths ($1-2\kms$) and low-velocity motions with respect
to the cloud gas, much smaller than those detected on the front side. 
This implies that the rear side lies close to the plane of the sky, 
in agreement with the $\sim 10\deg$ angle of the optical jet to the plane 
of the sky derived by Reipurth, Raga \& Heathcote (1992) and the 
above opening angle of $5\deg$. 
The velocity distribution undergoes a {\em positive gradient} with increasing 
distance to the source, up to offset $+50\arcsec$ along the jet. 
As the jet is blueshifted, this positive gradient is naturally explained 
if the tangential velocity is much larger than the component normal to the 
wall.

We present in Fig.~7 the CO intensity distribution as a function of velocity 
in cuts made parallel to the jet direction. Emission from the front cavity 
walls is detected from $-11\arcsec$ to $+11\arcsec$ from the jet axis. 
The location of the front and rear cavity walls is marked with thick white 
lines, in the cut made along the jet axis (offset $-0.4\arcsec$).  The 
velocity shift between the front and the rear wall 
increases strongly with distance $x$ to the source, which 
implies that the tangential velocity also increases. The 
velocity field of the front and rear walls is well described 
by a linear law $v \propto x$; the corresponding fits are displayed in 
Fig.~7. 

The maximum velocity ($\approx 1\kms$)
is observed ahead of the optical jet.  A small gas  
knot is detected at the tip of the jet between offsets $-45\arcsec$ 
(bow shock L) and $-60\arcsec$ (knot O) in the panel at $+1.4\kms$ 
(see Fig.~5). A
counterpart to this knot is detected in the redshifted gas, at $+9.5\kms$.  
Emission at a lower velocity ($+2.4\kms$) is detected upstream of the knot 
and off the jet axis (Fig.~5). 
The whole wall appears to be accelerated along the optical jet up to 
this knot, as can be seen on the cuts at $-11\arcsec$ and $+2.4\arcsec$ 
in Fig.~7). 
About $5\arcsec$ downstream of knot O the wings of bow shock P are detected 
(Fig.~8 in Rei97). Those wings are remarkably 
extended, $10\arcsec$, comparable to the width of the inner cavity ``hole''.
The molecular gas has probably been accelerated in the wings of 
bow shock P.

Near the source, the gas undergoes a strong acceleration, which 
terminates  at the location of optical knot F, at $27\arcsec$ from the 
source (Figs.~4-5; Rei97). A symmetrical feature displaying a strong gas 
acceleration is also detected , which extends up to $+15\arcsec$ East of the 
source. Both components are marked  by a  white dashed line in Fig.~7. 

We suggest that both features are tracing a recent ejection from the 
protostellar source, associated with knots E-F. This gas  component
extends from $-6\arcsec$ up to $+5\arcsec$ from the jet main axis (Fig.~7); 
hence the ejection takes place inside the large-scale cavity reported above.
Indeed, based on STIS spectroscopic data of the jet obtained with HST,
Raga et al. (2002) showed good evidence that optical knot F traces an internal 
working surface resulting from time variability in the ejection velocity, 
on a scale of $250\yr$. 

The velocity field appears much more complex (see panel at $-0.4\arcsec$ in 
Fig.~7). It can be fitted by a linear law between the source  and offset 
$\Delta \alpha= -20\arcsec$. One can note that  the linewidth gets broader 
with decreasing distance to knot F, in panels at $-3.2\arcsec$, $+2.4\arcsec$ 
and along the jet. 
Between $-25\arcsec$ and $-40\arcsec$, the CO emission covers a broader range 
of velocities. Such broadening could trace gas accelerated by knots 
E-F, or in previous events like bow shock L.

\section{The CO bullets}

Beyond the optical bow shock V, CR discovered three CO bullets
(hereafter referred to as Bullet 1, 2 and 3, in order of increasing
distance from the source) with separations of $\simeq 55\arcsec$. No optical
emission is associated with these bullets. However, they are perfectly
aligned along the direction defined by the optical jet.  
Whereas the optical knots P and V are tracing shocked material, 
the molecular bullets B1-2-3  have no heating source.
Based on the association of optical knots with some high-velocity bullets, 
CR proposed that the latter bullets were tracing past shock episodes of 
HH~111.

\subsection{The PdBI observations}

\begin{figure*}
  \begin{center}
    \leavevmode
    \includegraphics[width=6cm,angle=-90]{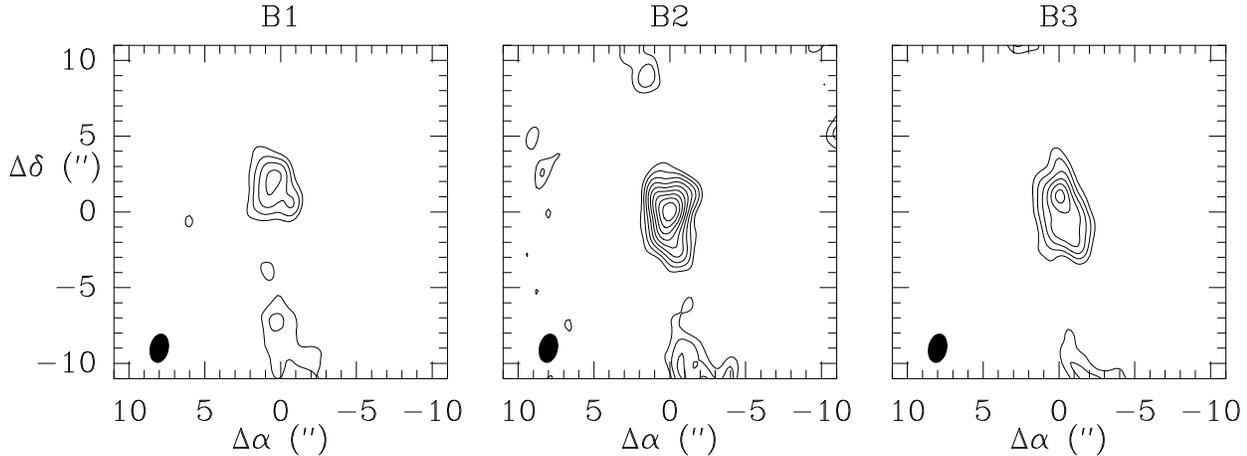}
\caption{
Interferometric maps of the velocity-integrated flux in the bullets B1, 
B2, B3 in the CO \jtwotoone\ transition. 
First contour interval are 0.10 and 0.04 Jy/beam respectively.
}
  \end{center}
\label{plot_bullet}
\end{figure*}

High-angular resolution maps have been obtained with the PdBI
interferometer in the CO \jtwotoone\ line (Fig.~8). 
The emission properties of all three bullets are very similar. 
The molecular emission from the bullets was marginally resolved by the IRAM 
30m-telescope; it allowed to estimate a size of $8\arcsec-10\arcsec$.
The much better angular resolution of the PdBI observations 
allows to resolve the emission.  Each bullet is elongated in the 
North-South direction; the large-scale distribution is well described by 
an ellipse with typical size (beam-deconvolved)
of $3\arcsec \times 7\arcsec$ ($2.0\times 4.8\times 10^{16}\cm$). 
We note that the transverse size of the bullets ($\simeq 10\arcsec$) matches 
reasonably well the inner diameter of the cavity around the jet.  

Each bullet presents a complex kinematical structure 
down to arcsec scale (Fig.~9). The line profiles
are characterized by a broad plateau, spanning $10-15\kms$, and a line core
of $5\kms$ HPFW. 
To study the gas kinematics in the bullets, we have considered separately
the spatial distributions of the blue and red wings (Fig.~10). In all 
three bullets, 
the blueshifted component appears to lie $\approx 1\arcsec$ (1 beam size) 
upstream of the redshifted component. It suggests  a side-ways 
expansion of the bullets, which now propagate in a low-density medium, and 
are oriented at only a small angle out of the plane of the sky.

\begin{figure*}
  \begin{center}
    \leavevmode
    \includegraphics[width=0.5\hsize,angle=-90]{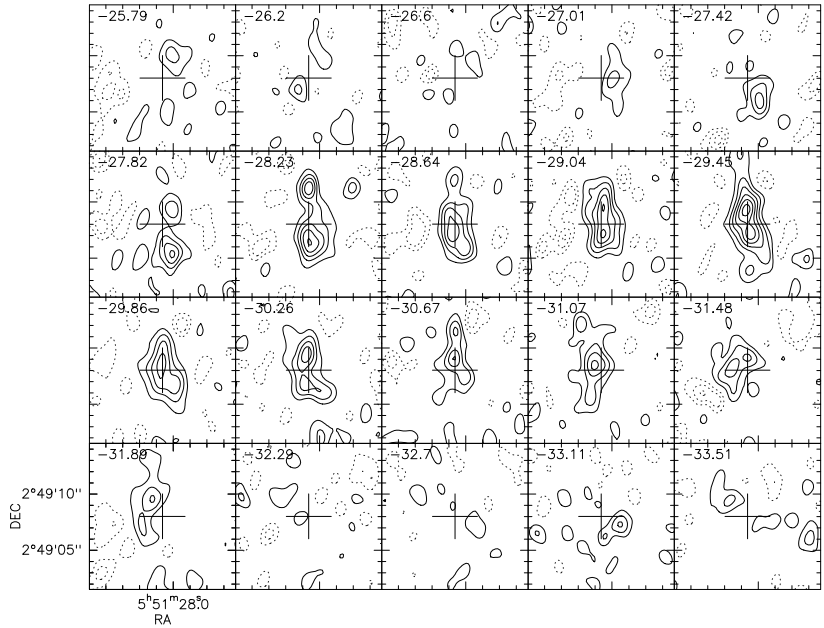}
\caption{Map of the \jtwotoone\ emission in B3, integrated in 
velocity intervals of $0.4\kms$. Contour interval is 0.2 Jy/beam (1.87~K).
}
  \end{center}
\label{b3_channelmap}
\end{figure*}

\begin{figure*}
  \begin{center}
    \leavevmode
    \includegraphics[width=0.5\hsize,angle=-90]{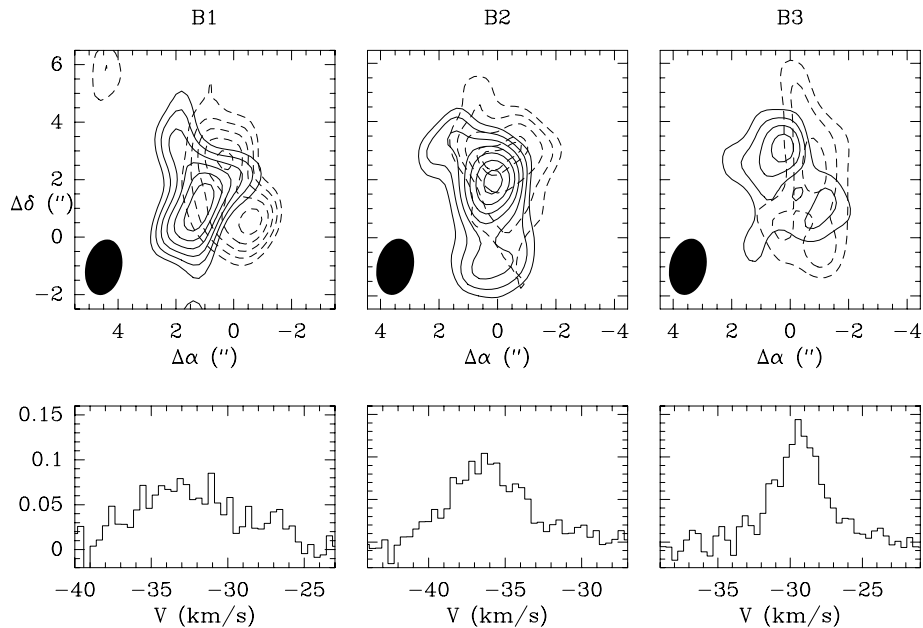}
\caption{({\em bottom})~Average spectrum of the bullets, in the region 
of intensity larger than 50\% the brightness peak. 
({\em top})~blueshifted (redshifted) emission in solid (dashed) contour.
First contour and contour interval are 0.10 and 0.04 Jy/beam, respectively.
}
  \end{center}
\label{plot_velmap}
\end{figure*}

\subsection{The physical conditions}

Hatchell, Fuller \& Ladd (1999) observed at the JCMT the CO \jfourtothree\ and 
\jthreetotwo\ emission near optical knot P, bow shock V and bullet B3. 
Under the hypothesis that the levels are populated at LTE and the lines 
are optically thin, they estimated a kinetic temperature of  $\simeq 100\K$ 
in the high-velocity ejecta. It is in bullet B3 that these authors derive 
the  highest temperature, around  $270\K$. This is rather unexpected as B3 is 
the most distant bullet from the source. We note that the transitions 
observed by the authors are not well suited to trace so high temperature 
gas. Higher-J transitions of CO are more sensitive probes of the kinetic 
temperature. 

\begin{table*}
\caption[]{Physical parameters of the CO bullets mapped with the  
PdBI.}
\begin{flushleft}
\begin{tabular}{c|c|l|l|l}
\hline\hline
Bullet  & Position     &  Size  & Mass & Density \\ 
        &  (J2000)     &  ($\arcsec$) & ($10^{-4}\msol$) & ($10^4\cmmt$)\\ 
\hline
B1      & 05:51:34.9, 02:48:51.1 &   $3.3 \times 4.6$  &  3.1  & 1.5 \\
B2      & 05:51:31.4, 02:48:57.7 & $3.3 \times 6.7$    &  2.1  & 0.59 \\
B3      & 05:51:29.0, 02:49:08.0 & $3.3 \times 6.7$    &  1.9  & 0.53 \\
\hline
\end{tabular}
\end{flushleft}

\end{table*}

We have carried out an LVG analysis of the CO line fluxes 
measured
towards B3, based on the data of CR and Hatchell et al. (1999). We 
have first corrected 
for the beam dilution, especially critical for the \jonetozero\ 
transition, adopting a bullet size of $10\arcsec$, in agreement with CR 
and our PdBI observations (see below). We adopted an average line width 
$\Delta v= 5.5\kms$, from the 30m and PdBI observations. We conclude 
that it is not possible to reproduce simultaneously all four transitions. 
The observed \jfourtothree\ flux exceeds by a large factor (about 4) the 
intensity 
predicted by the model. It is in fact the high flux measured in this transition
which is responsible for the high temperature derived by Hatchell et al. 
(1999). 
However, there is no evidence for internal shocks, 
which would heat the bullet material to such high temperatures. 
As discussed below, the cooling timescale
of this gas is very short, so that its temperature should decrease much below
the $100-150\K$ measured in the optical knots P-V.
We conclude that, most likely, the \jfourtothree\ 
line intensity was highly overestimated. As pointed out by Hatchell et al. 
(1999), the calibration of the 460~GHz receiver was highly uncertain at the 
time of the \jfourtothree\ observations. 

If we repeat the analysis considering only the first three CO transitions and 
taking into account
the uncertainty in the measured flux (about $10\%$), a good agreement
is obtained for a CO column density $\rm N(CO)=1.2\times 10^{16}\cmmd$. 
The best agreement is then obtained for 
temperatures in the range $35-45\K$ and \htwo\ densities in the range 
$5-8\times 10^3\cmmt$ (see below). 

Neufeld, Lepp \& Melnick (1993,1995) have studied the radiative cooling of
warm molecular gas $\rm T \geq 100\K$ in various astrophysical
environments. Applying their work to the case of the HH~111 bullets, we 
estimate a typical cooling time of $600\yr$ if the temperature of the 
bullets is as high as $100\K$. This is approximately the typical flight 
time between two subsequent molecular bullets ($500 \yr$). The distance 
from bullet V, the last one for which we observe evidence of a shock, 
to bullet B3 is $120\arcsec$, that needs about $1500\yr$ to be crossed, 
almost three times the CO cooling time.

We conclude that, although the temperature of the bullets inside 
the molecular cloud is about $100-150\K$, the molecular bullets 
B1-B2-B3 are most likely made of colder gas, at a temperature of typically 
$35-45\K$. Under such physical conditions ($n(\htwo)= 8000\cmmt$, 
$\rm N(CO)= 1.2\times 10^{16}\cmmd$), the \jtwotoone\ transition 
is optically thin $\tau\simeq 0.3$ and has an excitation temperature 
$T_{ex}\simeq 20\K$. This is  slightly larger than the value
derived by CR ($15\K$) although the other physical parameters 
(mass, mass transport rate, momentum) are essentially unchanged.

From the above timescale, we estimate a time-averaged mass transport 
of $4\times 10^{-7}\msol\yrmu$, whereas Hartigan, Morse \& Raymond (1995) 
derive a mass loss rate of $1.8\times 10^{-7}\msol\yrmu$. The similar mass 
transport rates support the view that the CO bullets are not made of ambient
material that has been accelerated to these very high velocities, but are
mostly composed of material ejected from the central engine. The momentum
carried away by each bullet amounts to $1.0\times 10^{-4}\msol\kms\yrmu$.

Raga et al. (1990, 2002) proposed that optical knots form from time variable 
gas 
ejections in the jet. Such mechanism leads to the formation of ``mini-bows''
along the jet. In this framework, the high-velocity CO bullets would 
trace the previously shocked molecular counterpart of the optical knots. It 
naturally 
explains the similar properties observed between the jet and the 
high-velocity  bullets (size, kinematics, momentum). The weak high-velocity
component detected between B1 and B2 (see Fig.~6 in CR) traces the molecular 
``intershock'' gas ejected from the central source, between two ``eruptions''.

\section{Gas entrainment mechanism}

\subsection{The cavity}
 
The kinematics of the outflowing gas is complex as it bears the signature 
of several ejection episodes in the optical jet,
associated with bow shocks at knots F and L (Rei97; Raga et al. 2002). 
We also find strong correlation between the low- and high- velocity
outflow wings and the jet. It is therefore unlikely 
that a single fit of any outflow model is  able to reproduce the kinematics of 
the gas observed along the whole flow length.  As the ejecta are located 
inside the cavity, close to the source (within $\approx 30\arcsec$), we have 
decided to concentrate on the global properties of the cavity, at larger 
distance from the source. 
Our interferometric data have revealed a large cylindrical cavity 
in which the gas is accelerated following a ``Hubble law'' 
$v \propto x$, distance along the jet axis.   

The formation of such large cavities is easily accounted for by 
the sweeping up of ambient material as a bow shock propagates 
into the parental cloud (RC93). 
In what follows, we show that that such gas entrainment models 
can indeed account for the molecular outflow  emission in HH~111. 
We first determine the parameters of the ``best'' model that fits the 
HH~111 molecular gas distribution. 

As shown by RC93, the global shape of the cavity swept-up 
by an internal bow shock depends on two parameters~: the density 
stratification of the ambient medium and the ratio between the distance of the 
working surface to the source $d$ and the standoff distance between the 
working surface and the bow shock apex $r_s$.

For a jet propagating into a medium of density $\rho \propto x^{\beta}$, 
$x$ measuring the distance from the powering source along the jet axis, 
the following relation holds between the jet cylindrical radius $r_c$ and 
$x$~: $r_c \propto x^{-\beta/2}$. 
As discussed above and in Sect.~4.3, the shape of the cavity {\em near} 
the source is suggestive of  $\beta= -1$. 

The standoff distance $r_s$ between the ejection point in 
the working surface and the bow shock apex is related to the jet radius 
via~: 
\begin{equation}
r_s = r_j \left(\frac{\pi\rho_j}{\Omega \rho_a}\right)^{1/2} 
\left(\frac{v_0}{v_j}\right)
\end{equation}
where $\Omega$ is the solid angle over which mass is ejected 
($\Omega \approx 2\pi\sqrt{2}$), $v_0$ and $v_j$ are the ejection and jet 
velocities, respectively (see Eq.~4 in RC93).

In the case of strong working surfaces such as those observed in HH~111, 
the ratio  between the jet and the ambient density $\rho_j/\rho_a$ is  
$\simeq 100$ (Raga \& Binette 1991) and $v_0\simeq v_j$ in the adiabatic 
case. Hence $r_s/r_j \simeq 6\times (v_0/v_j) \leq 6$ and 
$r_s\leq 9\arcsec$, with a typical jet radius of $1.5\arcsec$ (Rei97). 

A simple determination comes from the shape of the cavity computed 
in the case $\beta= -1$. The cylindrical radius is typically twice the 
standoff distance (see for example the top panel of Fig.~3); it is not very 
sensitive to the actual value of $\beta$ nor the distance of the bow shock 
to the source. At zeroth order, the diameter of the cavity determined in the 
outer part is $\approx 18\arcsec= 4 r_s$, hence $r_s\approx 4.5\arcsec$. 
This simple determination is consistent with the upper limit obtained 
previously. It is also consistent with numerical simulations, which 
yield $r_s/r_j \sim 3$. 

Since bow shock V lies at $148\arcsec$ from the source (Rei97), one has 
$d/r_s= 32$. We display in Fig.~6 the shape of the cavity swept-up by a bow 
shock as predicted for this set of parameters. 
This model fits nicely the walls of the cavity and the curvature of 
bow shock V.  The agreement is very satisfying at 
$\Delta \alpha \leq -50\arcsec$. 
We have fitted the inner walls of the cavity by straight lines (drawn in 
dashed dark contours in Fig.~6); the optical jet emission appears to fill 
exactly the cavity walls up to $\Delta \alpha= -50\arcsec$. Extrapolating the 
fit up to bow shock V, we find that the separation of the walls at that 
position is $\sim 10\arcsec$. It matches well the size of the bow shock 
in the $\rm H\alpha$ optical line (see top panel in Fig.~6), as expected 
for freely expanding shocked material.  

For values of $\beta$ of $-1$, the model predicts that the velocity 
vs position along the axis is almost exactly linear over the whole flow 
length, except close to the source. This is again consistent with the 
Position-Velocity diagram derived from our interferometric maps (see Fig.~7). 

The situation is more complex closer to the source~: molecular gas emission is 
still detected beyond the fit, so that the agreement between the fit and 
the actual shape of the cavity is not very good. Several factors may 
contribute to this. In particular, several ejections have taken place 
recently, which are now observed as knots E to L. Since this part of the 
cavity has experienced the crossing of two more bow shocks (P and L), 
the walls have been transferred additional momentum. This could contribute 
to our detection of entrained gas at further distances from the jet axis. 

These results differ somewhat from the previous conclusions of Nagar et al. 
(1997).
Based on their interferometric map of the $\rm CO \jonetozero$ line emission, 
these authors interpreted 
the velocity field and the spatial distribution of the molecular gas 
as evidence for an outflow entrained by a wide-angle radial wind, 
as modelled by Shu et al. (1991, 1995). The modelling of the 
Position-Velocity diagram of the 
emission along the jet is indeed in good agreement with the predictions of
radial-wind driven outflows (Li \& Shu 1996). 
As pointed out by Cabrit et al. (1997), a simple fit to 
the Position-Velocity distribution of the outflowing gas does not allow
to discriminate between jet-driven and wide-angle wind models, when 
the inclination with respect to the plane of the sky is low. 

We note that the physical conditions close to the source (less 
than $15\arcsec$)
play an important role in constraining the parameters of the wide-angle wind
fitted by Nagar et al. (1997). As can be seen on their Fig.~5, it appears that 
at larger distances the radius of the cavity is constant rather than 
parabolic-like, as claimed by the authors, and the match between the observed 
velocity profile and the fit is less satisfying. 

In our opinion, the strong ejection activity near the source prevents 
conclusions about the role of a wide-angle wind in driving molecular gas, 
as these ejecta are very efficient at transferring momentum to the ambient 
material.
Interferometric observations  at a higher angular resolution than the data 
presented here are required to investigate the relation the optical shocked 
knots hold to the molecular gas, and how the dynamics of the cavity is 
affected.  

However, the formation of the large-scale cavity observed up to bow shock P
is well explained
in the frame of the model of RC93. The jet entrains ambient molecular material
swept-up in the propagation of bow shocks, which disturb the ambient gas 
at large distances from the jet axis. 
Our observations indicate that bow shock V played a major role in shaping 
the global cavity.

\subsection{The hot gas component}

Our observations of the CO 7-6 emission bring evidence for hot outflowing gas
in the cavity which is missed by observations of the low-J transitions, 
due to the extremely low opacity of the latter. This is evidenced by the 
maximum radial velocity detected in the outflow wings in the \jseventosix\ 
and \jtwotoone\ transitions ($30\kms$ and $10\kms$, respectively). 
It is therefore necessary to take into account the thermal structure of the
outflow to discriminate between the various gas acceleration mechanisms 
that have been proposed.  

The hot gas component occupies probably a much smaller region 
than the low-velocity cold gas, with a typical size of the emitting region
of a few arcsec. The emission likely 
arises from shocked gas in the jet, a conclusion supported by 
the high velocities and temperature measured in this gas. High signal:noise
spectra obtained 
further downstream in the jet show a continuous velocity distribution 
between the molecular jet and the high-velocity wings, and support a close 
association between both phenomena. 
[CI] observations do not show any evidence of CO dissociation close to 
the source; hence,
the gas is probably accelerated in C-type, rather than J-type shocks, along 
the jet. 

Averaging the CO \jtwotoone\ signal over the cavity reveals a faint 
molecular jet signature at $\-60\kms$, consistent with the velocity 
of the molecular features (bullets) detected further downstream (Fig.~4).
The presence of such a hot component shows that 
there is no clear velocity separation between the outflowing gas and 
the jet itself (see Fig.~4, and lower panel of Fig.~1 in CR).
Interferometric observations with the SMA in the CO \jsixtofive\ transition
would allow to determine the location of the emission. Observations of the 
high-J CO lines with the Herschel Space Observatory  would  
allow to clarify the physical conditions in the outflowing gas.

\section{Conclusions}

We have presented single-dish and interferometric line observations of the 
HH~111 outflow and its driving source. 
The physical conditions of the protostellar core were determined from the 
emission of the millimeter lines of CO and its 
isotopomers, the millimeter lines of CS, all of them observed with the IRAM 
30m telescope,
and the CO \jseventosix\ line observed with the CSO. The molecular gas emission
reveals a small condensation of cold ($\rm T= 20-25\K$) and dense gas 
($n(\htwo)= 3\times 10^5\cmmt$). 

The low-velocity outflowing gas has been mapped with the IRAM Plateau de
Bure interferometer. It 
is distributed in a hollow cylinder that surrounds the optical jet.  
The formation of this cavity and its kinematics are well accounted 
for in the frame of the model of RC93. The jet entrains ambient molecular 
material swept-up in the propagation of bow shocks, which disturb the 
ambient gas at large distances from the jet axis. 
Our observations indicate that bow shock V played a major role in shaping 
the global cavity. The gas velocity is found to vary almost linearly with 
distance to the source over the whole flow length, except close to the source. 
The cavity has been expanding with a mean 
velocity of $4\kms$ on a timescale of $8700\yr$, similar to the 
dynamical age of the optical jet. The separation of the inner walls reaches 
$8-10\arcsec$ near
knots P and S, which corresponds to the transverse size of the wings in these 
bow shocks. 

The more recent ejections detected close to the source 
are still efficient at transferring momentum to the ambient material. 
Regions of strong gas acceleration are detected along the jet, which 
are correlated with the optical bow shocks at knots F and L.  

Observations of the high-excitation line CO \jseventosix\ reveal high-velocity
outflowing gas ($\rm v= 150\kms$ after deprojection) near the central source. 
This high-velocity component is hot ($T= 300-1000\K$) and fills only a small
fraction of the beam. We propose that this emission arises from shocked gas 
in the optical jet. 

Emission of the molecular bullets B1-2-3 was mapped with the IRAM PdBI in the 
CO \jtwotoone\ transition. The emission
is distributed in structures of $3\arcsec \times 7\arcsec$ flattened 
perpendicularly to the jet main axis. Their kinematics
is complex but an overall side-ways expanding motion is detected inside 
all of them.
Their physical properties have been reassessed; we find a low temperature 
for the bullets ($\sim 30\K$), in agreement with the cooling timescale 
for shocked gas initially heated at temperatures of $100\K$ or beyond, and 
cooling down while flying freely.  Their 
density lies now in the range $(0.5-1.0)\times 10^4\cmmt$. 
These molecular bullets probably form in the shocks resulting from 
time-variable ejections in the powering source, in the manner proposed 
by Raga et al. (2002) to account for the presence of mini bow shocks and knots 
in the optical jet. In such a scenario, the weak high-velocity molecular jet 
detected between the molecular bullets represents intershocked gas ejected 
from the central source. 

\acknowledgments

\end{document}